\documentclass[aps,prb,twocolumn,superscriptaddress,
groupedaddress,10pt]{revtex4}
\usepackage{amsmath}
\usepackage{amssymb}
\usepackage{graphicx}
\usepackage{epsfig}
\usepackage{dcolumn}
\usepackage{bm}
\usepackage{bm}
\usepackage{blindtext}
\usepackage{natbib}
\usepackage{siunitx} 

\usepackage[titletoc]{appendix}

\usepackage{simplewick}

\usepackage{bbm}
\makeatletter
\newcommand{\mathleft}{\@fleqntrue\@mathmargin0pt}
\newcommand{\mathcenter}{\@fleqnfalse}
\makeatother

\usepackage{afterpage}

\usepackage[usenames,dvipsnames]{xcolor}
\usepackage{amsthm}
\usepackage{booktabs}
\usepackage{hyperref}
\usepackage{tikz}
\usetikzlibrary{calc}
\usepackage{mathtools}

\def\nn{\nonumber}

\newcommand{\ket}[1]{| #1 \rangle}
\newcommand{\bra}[1]{\langle #1 |}

\begin{document}
\title{Nonrelativistic axion electrodynamics in $p+is$ superconductors}

\author{Chao Xu}
\affiliation{Kavli Institute for Theoretical Sciences, University of Chinese Academy of Sciences, Beijing 100190, China}
\affiliation{Department of Physics, University of California San Diego, 9500 Gilman Drive, La Jolla, California 92093, USA}

\author{Wang Yang}
\email{wang.yang@ubc.ca}
\affiliation{Department of Physics and Astronomy and Stewart Blusson Quantum Matter Institute, University of British Columbia, Vancouver, B.C., Canada, V6T 1Z1}

\begin{abstract}
In previous works, axion electrodynamics in three dimensional $p+is$ superconductors is discussed by borrowing the results from superconducting Dirac systems. 
However, in this work, based on a systematic path integral approach,
we show that the axion electrodynamics in $p+is$ superconductors exhibits a nonrelativistic form, which is different from the superconducting Dirac and Weyl systems.
More precisely, the induced electric field does not enter into the axion action, and gauge invariance is ensured by the combination of the electric potential and the Nambu-Goldstone phase mode. 
Furthermore, unlike the axion angle in the Dirac case which is equal to the difference between the superconducting phases on the two  Fermi surfaces of different helicities, 
the axion angle in the present case contains an additional sinusoidal term.
As applications of the derived nonrelativistic axion electrodynamics, physical effects related to vortex lines  and Witten effect are discussed.
Our work reveals the differences for axion electrodynamics between the relativistic and nonrelativistic systems.

\end{abstract}
\maketitle

\section{Introduction}

Spin triplet superconductivity and paired superfluidity have a complex spin-orbit entangled structure in the Cooper pair wavefunctions \cite{Leggett1975,Vollhardt1990,Volovik2003,Mackenzie2003,Leggett2006}, 
leading to exotic behaviors like  topological properties \cite{Hasan2010,Qi2011,Ando2015,Sato2016,Chiu2016,Sato2017}.
Another interesting type of Cooper pairing is the one with spontaneous time reversal symmetry breaking, which arises when two or more channels of pairing instabilities compete and coexist
\cite{Volovik1988,Volovik1989,Ivanov2001,Kopnin1991,Stone2006,Tewari2007,Chuanwei2008,Fu2008,Cheng2010,Qi2009,Wang2014,Wang2017,Yang2017,Xu2020,Lee2009,YiLi2013,Thomale2011,Platt2012,Khodas2012,Fernades2013,Hinojosa2014,Stanev2010,Lin2012,Marciani2013,Maiti2013,Ahn2014,Garaud2014,Maiti2015,Laughlin1998,Senthil1999,Horovitz2003,Hu2008,Sato2010,Black2012,Chubukov2012,Wang2012,Kiesel2013,Liu2013,Black2014,Wu2010}.
A mixture of triplet and singlet Cooper pairings which breaks time reversal symmetry has been theoretically studied in different contexts, 
showing exotic properties including nontrivial bulk electromagnetic and gravitational responses \cite{Ryu2012,Qi2013}, 
quantized surface thermal Hall effects \cite{Read2000,Wang2011,Ryu2012,Stone2012}, 
chiral Majorana fermions propagating along the magnetic domain wall on the surface \cite{Yang2017},
and high order topology \cite{Roy2020}.
Recently, there has been experimental evidence of triplet pairing gap functions with spontaneous time reversal symmetry breaking  in real materials \cite{Shang2018,Sundar2019}.

Axion as an elementary particle was proposed in the high energy context more than four decades ago \cite{Peccei1977,Weinberg1978,Wilczek1978},
which has been considered as a candidate for dark matter and dark energy,
though its existence still remains inconclusive.
On the other hand, the dynamical axion field has been proposed to exist in topological systems as a condensed matter realization of axions \cite{Li2010,Wang2011,Qi2013}.
The coupling between the axion angle $\theta$ and the electromagnetic field is of the form $\int d^4x \theta \vec{E}\cdot \vec{B}$ (up to an overall numerical constant factor),
resulting in various magnetoelectric effects,
where $\vec{E}$ and $\vec{B}$ are the electric and magnetic fields, respectively. 
In particular, the three-dimensional (3D) $p+is$ superconductor has been considered as a superconducting platform which hosts axion field \cite{Qi2013,Goswami2014,Shiozaki2014,Stone2016},
where the triplet pairing component is invariant under spin-orbit coupled rotations analogous to the pairing of the $^3$He-B superfluid.

In this work, we perform a systematic derivation of the coupling between the axion angle and the electromagnetic fields in 3D $p+is$ superconductors based on a  path integral approach,
including contributions from  both the orbital and Zeeman  channels.
The axion electrodynamics  in $p+is$ superconductors has been discussed in previous works by borrowing the results from superconducting Dirac systems \cite{Qi2013}. 
However, we find that there are crucial differences in axion electrodynamics between the $p+is$ superconductors and the superconducting Dirac/Weyl cases.
First, the axion action in $p+is$ superconductors is not of the $\vec{E}\cdot \vec{B}$ form, 
but is $\nabla(\phi+\hbar\partial_t \Phi/e)\cdot \vec{B}$,
where $\phi$ and $\Phi$ are the electric potential and the superfluid phase, respectively. 
In particular, the induced electric field $\partial_t \vec{A}/c$ does not appear in the action,
reflecting the fact that the axion electrodynamics in $p+is$ superconductors is nonrelativistic in nature.
Second, the axion angle $\Theta_{\text{ax}}$ in the $p+is$ superconductors is not just $\theta_{\text{ax}}$, 
defined as the phase difference between the superconducting phases on the two  Fermi surfaces of different helicities.  
In addition to $\theta_{\text{ax}}$,  $\Theta_{\text{ax}}$ also acquires a sinusoidal term $\sin(\theta_{\text{ax}})$.
As applications of the derived  nonrelativistic axion action,
we discuss the inflow currents to a vortex line of $\Theta_{\text{ax}}$ on the surface of the superconducting bulk, as well as the Witten effect. 

The rest of the paper is organized as follows.
In Sec. \ref{sec:Ham}, we introduce the model Hamiltonian for the $p+is$ superconductors,
briefly describe the axion action, 
and present a quick real space calculation of the orbital part of the axion action. 
Sec. \ref{sec:effective_action} presents a reformulation of the problem in the path integral framework. 
Sec. \ref{sec:orbital} and Sec. \ref{sec:Zeeman} are devoted to path integral derivations of the axion electrodynamics in the orbital and spin spaces, respectively. 
In Sec. \ref{sec:physical}, vortex lines and Witten effect are discussed. 
Finally in Sec. \ref{sec:conclusion}, we briefly summarize the main results of the paper.

\section{Model Hamiltonian and description of nonrelativistic axion action}
\label{sec:Ham}

\subsection{Model Hamiltonian}
\label{subsec:Ham}

We consider a superconducting 3D centrosymmetric electronic system which exhibits a mixture of singlet and triplet pairing symmetries.
The band dispersion is 
\begin{eqnarray}
\xi_\alpha(\vec{k})=\frac{\hbar^2}{2m}k^2-\epsilon_F,
\end{eqnarray}
in which $\epsilon_F=\frac{\hbar^2}{2m}k_f^2$ is the Fermi energy where $k_f$ is the Fermi wavevector,
and $\alpha=\uparrow,\downarrow$ is the spin index.
The pairing Hamiltonians $\hat{P}_s(\vec{k})$ and $\hat{P}_p(\vec{k})$ for the $s$-wave and $^3$He-B like $p$-wave pairing gap functions are defined as 
\begin{eqnarray}
\hat{P}^\dagger_s(\vec{k})&=& (i\sigma_2)_{\alpha\beta} c_\alpha^\dagger(\vec{k}) c_\beta^\dagger(-\vec{k}),\nn\\
\hat{P}^\dagger_p(\vec{k})&=&\frac{1}{k_f} k_j(i\sigma_j\sigma_2)_{\alpha\beta} c_\alpha^\dagger(\vec{k}) c_\beta^\dagger(-\vec{k}),
\label{eq:Define_P}
\end{eqnarray}
respectively, 
in which: $\alpha,\beta$ are spin indices; 
$c_\alpha^\dagger(\vec{k})$ is the electron creation operator with momentum $\vec{k}$ and spin $\alpha$;   
$\sigma_j$'s ($j=1,2,3$) are the three Pauli matrices in spin space;
and repeated indices imply summations.
We note that $\hat{P}_p(\vec{k})$ is invariant under spin-orbit coupled SO(3) rotations,
which has the same form as the pairing in the $^3$He-B superfluid \cite{Leggett1975}. 

The pattern of the mixed-parity pairing gap function can be determined by a Ginzburg-Landau free energy analysis.
Keeping up to quartic terms and neglecting terms involving temporal and spatial derivatives, 
the most general form of the free energy invariant under both time reversal ($\mathcal{T}$) and inversion ($\mathcal{P}$) symmetries is given by
\begin{eqnarray}
F&=& -\alpha_s \Delta_s^{*}\Delta_s-\alpha_p \Delta_p^{*} \Delta_p+\beta_s |\Delta_s|^4+\beta_p |\Delta_p|^4\nn\\
&&+\gamma_1 |\Delta_p|^2  |\Delta_s|^2 +\gamma_2 (\Delta_p^{*} \Delta_p^{*} \Delta_s \Delta_s+c.c),
\label{eq:GL}
\end{eqnarray}
in which $\Delta_\lambda$ ($\lambda=s,p$) are the pairing gap functions in the $\lambda$-channel.
When the instabilities in the $s$- and $p$-wave channels coexist, 
both $\alpha_s$ and $\alpha_p$ are negative.
At tree level, the coefficients $\beta_\lambda,\gamma_j$ ($\lambda=s,p$, $j=1,2$)
are all determined by the electronic band structure, independent of the interactions.
In particular, close to the superconducting transition point, $\gamma_2=\frac{5\zeta(3)}{8\pi^2T_c^2} N_F$ 
is generically positive \cite{Wang2017}, where $\zeta(z)$ ($z\in \mathbb{C}$) is  the Riemann zeta function;
$T_c$ is the superconducting transition temperature (where for simplicity, a degenerate transition temperature for both $s$- and $p$-wave channels is assumed); 
and $N_F=\frac{mk_f}{2\pi^2\hbar^2}$ is the density of states at the Fermi level for a single spin component.
An important implication of a positive $\gamma_2$ is that a relative $\pi/2$ phase difference between $\Delta_s$ and $\Delta_p$ is energetically favorable as can be readily seen from Eq. (\ref{eq:GL}), 
leading to a superconducting pairing of the $p\pm is$ form.
Notice that the $p\pm is$ pairing spontaneously breaks both time reversal and inversion symmetries, but remains invariant under the combined $\mathcal{PT}$-operation up to an overall gauge transformation \cite{Yang2017}.
We  note that in addition to the intrinsic $p\pm is$ superconductors, the $p\pm is$ pairing symmetry can also be realized via proximity effects, as discussed in Ref. \onlinecite{Yang2017}.

In the remaining parts of this article,  $|\Delta_\lambda|$ is denoted as $\Delta_\lambda$ for short, 
and the $\pi/2$ superconducting phase difference will be explicitly displayed by considering the $p+is$ pairing Hamiltonian $\Delta_p\hat{P}^\dagger_p(\vec{k})+i\Delta_s\hat{P}^\dagger_s$.
In the Bogoliubov-de-Gennes (BdG) formalism, the mean field Hamiltonian acquires the form
\begin{eqnarray}
H_{\text{BdG}}=\frac{1}{2}\sum_{\vec{k}}\Psi^\dagger(\vec{k})H(\vec{k})\Psi(\vec{k}),
\end{eqnarray}
in which 
\begin{eqnarray}
\Psi^\dagger(\vec{k})=(c^\dagger_\uparrow(\vec{k})~ c^\dagger_\downarrow(\vec{k}) ~c_\uparrow(-\vec{k})~c_\downarrow(-\vec{k})),
\label{eq:Psi_k}
\end{eqnarray}
and the matrix kernel $H(\vec{k})$ is 
\begin{eqnarray}
H(\vec{k})=(\frac{\hbar^2}{2m}k^2-\epsilon_F)\gamma^0+\frac{\Delta_p}{k_f}\vec{k}\cdot\vec{\gamma}+\Delta_s \gamma^4,
\label{eq:Hk}
\end{eqnarray}
in which $\vec{\gamma}=(\gamma^1,\gamma^2,\gamma^3)$, and the matrices $\gamma^\mu$ ($\mu=0,1,2,3,4$) are defined as
\begin{flalign}
\gamma^0=\sigma_0\tau_3,\gamma^1=-\sigma_3\tau_1,\gamma^2=-\sigma_0\tau_2,
\gamma^3=\sigma_1\tau_1,\gamma^4=\sigma_2\tau_1,
\end{flalign}
where $\tau^i$ ($i=1,2,3$) are the Pauli matrices in the Nambu space, 
and $\sigma_0$  and $\tau_0$ denote the $2\times 2$ identity matrices in the spin and Nambu spaces, respectively.
It can be straightforwardly verified that the five gamma-matrices $\gamma^\mu$ ($\mu=0,1,2,3,4$) satisfy the anticommutation relations
\begin{eqnarray}
\{\gamma^\mu,\gamma^\nu\}=2\delta_{\mu\nu}.
\label{eq:anti_comm}
\end{eqnarray}

Later we will also consider spatially and temporally varying pairing gap functions $\Delta_\lambda(\vec{r},t)$ ($\lambda=s,p$).
Including the minimal coupling to electromagnetic potentials $\{A^\mu\}_{\mu=0,1,2,3}$,
 the Hamiltonian in real space becomes
\begin{eqnarray}
H_{\text{BdG}}=\frac{1}{2}\int d^3r
\sum_{\vec{r}}\Psi^\dagger(\vec{r})
\hat{T}^\rho\gamma^\rho
\Psi(\vec{r}),
\label{eq:BdG_realspace}
\end{eqnarray}
in which the summation is over $\rho=0,1,2,3,4$;
 $\Psi^\dagger(\vec{r})$ is the Fourier transform of $\Psi^\dagger(\vec{k})$; and
\begin{eqnarray}
\hat{T}^0&=&\frac{\hbar^2}{2m}(-i\nabla+\frac{e}{c}\vec{A})^2-\mu(\vec{r}),\nn\\
\hat{T}^j&=& \frac{1}{2k_f}\{ \Delta_p(\vec{r},t),-i\partial_j \}\gamma^j,~(j=1,2,3),\nn\\
\hat{T}^4&=&\Delta_s(\vec{r},t),
\label{eq:def_T}
\end{eqnarray}
where the convention for the electric charge is $e>0$.
In addition to Eq. (\ref{eq:BdG_realspace}),
there is also a Zeeman term in the presence of a magnetic field, i.e.,
\begin{eqnarray}
H_{\text{ZM}}=\frac{1}{2}g\mu_B\int d^3r
\sum_{\vec{r}}\Psi^\dagger(\vec{r})
[-\frac{i}{2}\epsilon_{ijk}B_i\gamma^j\gamma^k]
\Psi(\vec{r}),
\end{eqnarray}
in which $g$ is the Land\'e factor, $\mu_B=\frac{e\hbar}{2mc}$ is the Bohr magneton, 
 $\vec{B}=\frac{1}{c}\nabla\times \vec{A}$ is the magnetic field,
 and the $1/2$ factor in the front comes from the spin-$1/2$ nature of the electrons.
 
\subsection{Description of nonrelativistic axion action}
\label{subsec:description}

In Sec. \ref{sec:orbital} and Sec. \ref{sec:Zeeman}, we are going to derive the axion action based on a path integral approach. 
Here we briefly describe the main results.
Physical effects of the axion action will be discussed in Sec. \ref{sec:physical}.

The axion action in the $p\pm is$ superconductors is derived as
\begin{eqnarray}
S_{\text{ax}}=\frac{\alpha}{24\pi^2}\int d^4x \Theta_{\text{ax}}\nabla (\phi+\frac{\hbar}{e}\partial_t \Phi)\cdot \vec{B},
\label{eq:Full_axion_action_1}
\end{eqnarray}
in which 
$\alpha$ is the fine structure constant;
the axion angle $\Theta_{\text{ax}}=\Theta_{\text{ax}}^o+\Theta_{\text{ax}}^Z$ is 
\begin{eqnarray}
\Theta_{\text{ax}}=(1+g)(\theta_{\text{ax}}-\sin\theta_{\text{ax}}),
\label{eq:def_Theta}
\end{eqnarray}
where 
$g$ is the Land\'e factor,
$\theta_{\text{ax}}=\pi-2\arctan(\Delta_s/\Delta_p)$,
$\Theta_{\text{ax}}^o=\theta_{\text{ax}}-\sin\theta_{\text{ax}}$ is the orbital contribution,
and $\Theta_{\text{ax}}^Z=g\Theta_{\text{ax}}^o$ is the contribution from the spin channel.

There are notable differences between Eq. (\ref{eq:Full_axion_action_1}) and the conventional axion action in the relativistic case. 
In a relativistic system, the axion action acquires the form $\sim\int d^4x  \theta_{\text{ax}} \vec{E}\cdot \vec{B}$ where $\vec{E}=-\nabla\phi-\partial_t \vec{A}$.
On the other hand, Eq. (\ref{eq:Full_axion_action_1}) only contains the $\nabla\phi\cdot \vec{B}$ term,
and gauge invariance is ensured by adding $\frac{\hbar}{e}\partial_t\Phi$ to $\nabla\phi$.
Hence, in contrast to the relativistic case, the induced electric field $-\partial_t \vec{A}$ does not appear in Eq. (\ref{eq:Full_axion_action_1}).
The essential reason for such difference is a lack of Lorentz symmetry in the $p+is$ case.
We note that on a technical level, the band dispersion in the normal metal phase of $p+is$ superconductor does not contain negative energy states. 
Therefore, unlike the relativistic case, there is no inter-band transition (from negative to positive energy bands) in the $p+is$ superconducting case,
which leads to the different behaviors between the two situations.

In addition to the missing of $-\partial_t \vec{A}$, the axion angle in Eq. (\ref{eq:Full_axion_action_1}) contains an additional  sinusoidal term compared with the conventional relativistic Dirac case \cite{Stone2016}. 
Notice that the pairings on the two degenerate Fermi surfaces of different helicities (i.e., $\vec{k}\cdot\vec{\sigma}=\pm 1$) are $|\Delta|e^{\pm i\theta_{\text{ax}}}$ where $|\Delta|=\sqrt{\Delta_s^2+\Delta_p^2}$.
However, unlike the relativistic case, the axion angle $\Theta^o_{\text{ax}}$ is not just  $\theta_{\text{ax}}$ which is the difference between the superconducting phases on the two Fermi surfaces, but also contains $\sin(\theta_{\text{ax}})$.

\subsection{Real space picture of transverse current response}
\label{subsec:real_space} 
 
The calculations in Sec. \ref{sec:orbital} and Sec. \ref{sec:Zeeman} are rather formal and technical. 
In this subsection, we briefly present an alternative calculation in real space, which is more intuitive.
We emphasize that the calculation in this subsection only concerns with the $\nabla \phi$ contribution to the axion action in the orbital channel,
and rigorous treatments for both $\nabla \phi$ and $\partial_t \vec{A}$ as well as the Zeeman channel are included in Sec. \ref{sec:orbital} and Sec. \ref{sec:Zeeman}. 
Details of the real space calculations are included in Appendix \ref{sec:transverse_current}.

Focusing on the orbital part, the axion contribution to the electric current $\vec{j}$ can be obtained from $\vec{j}=c\frac{\partial S_{ax}}{\partial \vec{A}}$ as
\begin{eqnarray}
\vec{j}=-\frac{\alpha c}{24\pi^2} \nabla (\phi+\frac{\hbar}{e}\partial_t \Phi)\times \nabla\Theta^o_{\text{ax}},
\label{eq:orbital_j}
\end{eqnarray}
where $c$ is the light velocity.
From Eq. (\ref{eq:orbital_j}), it can be seen that  a transverse supercurrent can be induced by static electric field and spatial inhomogeneity of $\Theta^o_{\text{ax}}$.
We will derive Eq. (\ref{eq:orbital_j}) based on a real space approach, by assuming nonvanishing $\nabla \phi$ and $\nabla \Delta_s$.

In BdG form, the matrix kernel of the operator of the electric current density $j_i(\vec{x})$ ($i=1,2,3$) at position $\vec{x}$ is
\begin{flalign}
\hat{j}_i(\vec{x})= -\frac{e\hbar}{4m}\big[ \delta(\hat{\vec{r}}-\vec{x})(-i\nabla_{\vec{x}})+(-i\nabla_{\vec{x}})\delta(\hat{\vec{r}}-\vec{x})\big] \sigma_0\tau_0,
\label{eq:def_j_bdg}
\end{flalign}
in which $\hat{\vec{r}}$ is the coordinate operator.
The expectation value of $\hat{j}_i(\vec{x})$ is
\begin{eqnarray}
\left<\hat{j}_i(\vec{x})\right>=\text{Tr}(\hat{j}_i(\vec{x})\hat{G}),
\label{eq:j_expectation}
\end{eqnarray}
in which the Green's function $\hat{G}$ is
\begin{eqnarray}
\hat{G}=\frac{1}{-\partial_\tau-H},
\end{eqnarray}
where $\tau$ is the imaginary time, and $H$ is the matrix kernel of the Hamiltonian $H_{\text{BdG}}$
in the presence of a spatially varying electric potential $\phi(\vec{r})$ and $s$-wave pairing gap function $\Delta_s(\vec{r})$.
We emphasize that the symbol ``$\text{Tr}$" denotes the trace operation of an operator, which, in addition to the trace of the matrix structure in the spin and Nambu spaces, 
also involves the integral over spatial coordinates.
In what follows, we use ``$\text{tr}$" to indicate the trace which is only taken over the $4\times 4$ matrix structure. 

The Green's function $\hat{G}$ can be rewritten as
\begin{eqnarray}
\hat{G}=(-)\frac{1}{-\partial_\tau^2+H^2}(-\partial_\tau+H).
\label{eq:G_square}
\end{eqnarray}
The Hamiltonian squared $H^2$ can be separated as
\begin{eqnarray}
H^2=H_0^2+Q,
\label{eq:H_2}
\end{eqnarray}
in which
\begin{eqnarray}
H_0^2&=&\hat{T}^\rho \hat{T}^\rho,\nn\\
Q&=&\sum_{0\leq\rho<\sigma\leq4} [\hat{T}^\rho,\hat{T}^\sigma]\gamma^\rho\gamma^\sigma,
\label{eq:H0_2}
\end{eqnarray}
where $\hat{T}^\rho$ is defined in Eq. (\ref{eq:def_T})
and the anticommutation relations Eq. (\ref{eq:anti_comm}) is used.

Expanding $\left<\hat{j}_i(\vec{x})\right>$ in powers of $Q$, we obtain
\begin{eqnarray}
\left<\hat{j}_i(\vec{x})\right>&=&\sum_{n=0}^\infty (-)^{n+1}\text{Tr} \big[ \hat{j}_i(\vec{x}) 
(\frac{1}{-\partial_\tau^2+H_0^2}Q)^n\nn\\
&&\times\frac{1}{-\partial_\tau^2+H_0^2}(-\partial_\tau+H)
 \big].
 \label{eq:j_expansion}
\end{eqnarray}
Both the $n=0$ and $n=1$ terms vanish in Eq. (\ref{eq:j_expansion}).
The lead nonvanishing term comes from $n=2$, which gives (for details, see Appendix \ref{sec:transverse_current})
\begin{eqnarray}
&\left<\hat{j}_i^{(2)}(\vec{x})\right>=2(\frac{\Delta_p}{k_f})^3 \text{Tr}\big[
\hat{j}_i(\vec{x}) \frac{1}{-\partial_\tau^2+H_0^2}\partial_j\mu\frac{1}{-\partial_\tau^2+H_0^2}\partial_k \Delta_s \nn\\
&\times\frac{1}{-\partial_\tau^2+H_0^2}(-i\partial_i)
\big] \text{tr}(\gamma^0\gamma^j\gamma^k\gamma^4\gamma^i).
\label{eq:j_2_a_2}
\end{eqnarray}
Further evaluations of Eq. (\ref{eq:j_2_a_2}) give (for details, see Appendix \ref{sec:transverse_current})
\begin{eqnarray}
j_i(\vec{x})=e^2D(\Delta_s,\Delta_p)\epsilon_{ijk}\partial_j\phi(\vec{x})\partial_k\Delta_s(\vec{x}),
\label{eq:j_i_expression_2}
\end{eqnarray}
where
\begin{eqnarray}
D(\Delta_s,\Delta_p)=\frac{1}{6\pi^2\hbar} \frac{\Delta_p^3}{(\Delta_p^2+\Delta_s^2)^2}.
\label{eq:value_D_2}
\end{eqnarray}

We note that $D(\Delta_s,\Delta_p)$ can be written as a derivative,
\begin{eqnarray}
D(\Delta_s,\Delta_p)=-\frac{1}{24\pi^2\hbar}\frac{\partial \Theta^o_{\text{ax}}}{\partial \Delta_s},
\label{eq:def_Theta_2}
\end{eqnarray}
Imposing the boundary condition $\Theta^o_{\text{ax}}=0$ for a pure $s$-wave superconductor (i.e., $\Delta_s\gg\Delta_p$),
$\Theta^o_{\text{ax}}$ is determined to be
\begin{eqnarray}
\Theta^o_{\text{ax}}(\Delta_s,\Delta_p)&=&24\pi^2\int_{\Delta_s}^\infty dx D(x,\Delta_p) \nn\\
&=&\pi-\frac{2\Delta_s\Delta_p}{\Delta_s^2+\Delta_p^2} -2\arctan (\frac{\Delta_s}{\Delta_p}).
\end{eqnarray}
Therefore, defining
\begin{eqnarray}
\theta_{\text{ax}}=\pi-2\arctan(\frac{\Delta_s}{\Delta_p}),
\end{eqnarray}
$\Theta^o_{\text{ax}}$ can be written as
\begin{eqnarray}
\Theta^o_{\text{ax}}=\theta_{\text{ax}}-\sin(\theta_{\text{ax}}).
\label{eq:Theta_ax_expression}
\end{eqnarray}
Plugging Eq. (\ref{eq:Theta_ax_expression}) into Eq. (\ref{eq:j_i_expression_2}), we obtain Eq. (\ref{eq:orbital_j}).
This provides a real space derivation for the transverse supercurrent, 
which indirectly gives the orbital part of the axion action via the relation $\vec{j}=c\frac{\partial S_{ax}}{\partial \vec{A}}$.

\section{Path integral formulation}
\label{sec:effective_action}


In this section, we formulate the systematic path integral approach to the axion action in $p+is $ superconductors.
We assume the four-fermion interaction to be of the form
\begin{flalign}
H_{\text{int}}=-g_s\frac{1}{L^3}\sum_{\vec{k}} P_s^\dagger(\vec{k})P_s(\vec{k})-g_p\frac{1}{L^3}\sum_{\vec{k}} P_p^\dagger(\vec{k})P_p(\vec{k}),
\end{flalign}
in which $L$ is the linear size of the system in space and $g_\lambda>0$ ($\lambda=s,p$) are coupling constants,
and $P^\dagger_\lambda(\vec{k})$'s ($\lambda=s,p$) are defined in Eq. (\ref{eq:Define_P}).
After performing a Hubbard-Stratonovich transformation, 
the partition function in the imaginary time formalism can be written as 
\begin{eqnarray}
\mathcal{Z}=\int D[c^\dagger,c]D[\Delta_s^*,\Delta_s]D[\Delta_p^*,\Delta_p]e^{-\mathcal{S}},
\end{eqnarray}
in which
\begin{eqnarray}
\mathcal{S}=S_{\Delta}+S_f,
\end{eqnarray}
where
\begin{eqnarray}
S_{\Delta}=\sum_{\lambda=s,p}\frac{1}{g_\lambda}\int d\tau d^3r |\Delta_\lambda|^2,
\end{eqnarray}
and the fermionic part $S_f$ is
\begin{eqnarray}
S_f=\frac{1}{2}\int d\tau d^3r \Psi^\dagger(\tau,\vec{r})  (\partial_\tau+H_f) \Psi(\tau,\vec{r}),
\label{eq:Sf}
\end{eqnarray}
in which $\Psi^\dagger(\tau,\vec{r})$ is a set of  Grassmann numbers  defined through the Fourier transform of $\Psi^\dagger(\vec{k})$ in Eq. (\ref{eq:Psi_k}),
and $H_f$ is given by
\begin{eqnarray}
H_{f}^{11}&=&\frac{1}{2m} (-i\hbar\nabla +\frac{e}{c}\vec{A})^2-\epsilon_F-ie\phi+g\mu_B \vec{B}\cdot \vec{\sigma},\nn\\
H_{f}^{12}&=&[\frac{1}{2k_f}e^{-i\Phi}\{\Delta_p,-i\nabla\}e^{-i\Phi}\cdot \vec{\sigma}+i\Delta_se^{-i(2\Phi+\Phi_l)}] i\sigma_2,\nn\\
H_{f}^{21}&=&H_{f(1,2)}^\dagger,\nn\\
H_{f}^{22}&=&-[\frac{1}{2m} (-i\hbar\nabla -\frac{e}{c}\vec{A})^2-\epsilon_F-ie\phi+g\mu_B \vec{B}\cdot \vec{\sigma}^T],\nn\\
\end{eqnarray}
in which $H_{f}^{ij}$ ($i,j=1,2$) is the $(i,j)$-block of $H_f$;
$i\phi$ is the electric potential in imaginary time;
$\{A,B\}=AB+BA$ denotes the anticommutator of  the operators $A$ and $B$;
and $\Phi$ and $\Phi_l$ are the superconducting phase mode and the Leggett mode, respectively.

The phase mode $\Phi$ can be absorbed into electromagnetic potentials by performing a gauge transformation \cite{} through the following replacements:
\begin{eqnarray}
-ie\phi&\rightarrow  & \phi^\prime =-ie\phi+\partial_\tau \Phi,\nn\\
\frac{e}{c}\vec{A}& \rightarrow & \vec{A}^\prime=\frac{e}{ c}\vec{A}-i\hbar \nabla \Phi.
\label{eq:prime_A}
\end{eqnarray}
Assuming a background $\Delta_\lambda$ ($\lambda=s,p$) and including small fluctuations of the different modes, $H_f$ becomes
\begin{eqnarray}
H_f&=&H_{f0}+\Delta H_f,\nn
\end{eqnarray}
in which $H_{f0}$ is simply Eq. \ref{eq:Hk}, and 
\begin{eqnarray}
\Delta H_f&=&\Delta H_A^{(1)}+\Delta H_A^{(2)}+\Delta H_\phi+\Delta H_Z\nn\\
&&+\Delta H_p+\Delta H_s+\Delta H_{l},
\end{eqnarray}
where
\begin{eqnarray}
\Delta H_A^{(1)}&=&\frac{\hbar}{2m} \{\vec{A}^\prime(\tau,\vec{r}), -i\nabla\}\sigma_0\tau_0,\nn\\
\Delta H_A^{(2)}&=&\frac{1}{2m}[\vec{A}^\prime(\tau,\vec{r})]^2\gamma^0,\nn\\
\Delta H_\phi&=&\phi^\prime(\tau,\vec{r})\gamma^0,\nn\\
\Delta H_Z&=&-\frac{i}{4}g\mu_B \epsilon_{ijk}B_i\gamma^j\gamma^k,\nn\\
\Delta H_p&=&\frac{1}{2k_f} \{ \delta \Delta_p(\tau,\vec{r}),-i\nabla\} \cdot \vec{\gamma},\nn\\
\Delta H_s&=& \delta \Delta_s(\tau,\vec{r}) \gamma^4,\nn\\
\Delta H_l&=&\Delta_s \delta\Phi_l(\tau,\vec{r}) \sigma_2\tau_2.
\end{eqnarray}
We note that in momentum space, $\Delta H_A^{(1)}(\vec{q})=\frac{\hbar}{2m}(2\vec{k}+\vec{q})\cdot \vec{A}^\prime(\vec{q}) $ and $\Delta H_p(\vec{q})=\frac{1}{k_f}(\vec{k}+\vec{q}/2)\cdot \delta \Delta_p(\vec{q})$. 

Since the fermionic quasiparticles  are fully gapped, the action for the collective bosonic degrees of freedom  can be obtained by integrating over the fermions, resulting in $\frac{1}{2}\text{Trln}(-\partial_\tau-H_f) $.
In what follows, we will only consider the axion terms.
They arise in the third order terms in the Trln-expansion, i.e.,
\begin{eqnarray}
S^{(3)}_f=-\frac{1}{6}\text{Tr}[(\mathcal{G}_0\Delta H_{f})^3],
\label{eq:S3}
\end{eqnarray}
in which $\mathcal{G}_0=(\partial_\tau+H_0)^{-1}$.
Here we note that as discussed in Sec. \ref{sec:orbital} and Sec. \ref{sec:Zeeman}, in addition to the axion terms,
there are other nonvanishing terms in $S_f^{(3)}$ which involve two spacetime derivatives,
as a consequence of a lack of Lorentz symmetry.
We do not explicitly calculate these additional terms since the calculations are very cumbersome. 
A list of such terms based on a symmetry analysis 
is included in Appendix \ref{app:nonaxion}.

\begin{figure*}[htbp]
\centering
\includegraphics[width=16cm]{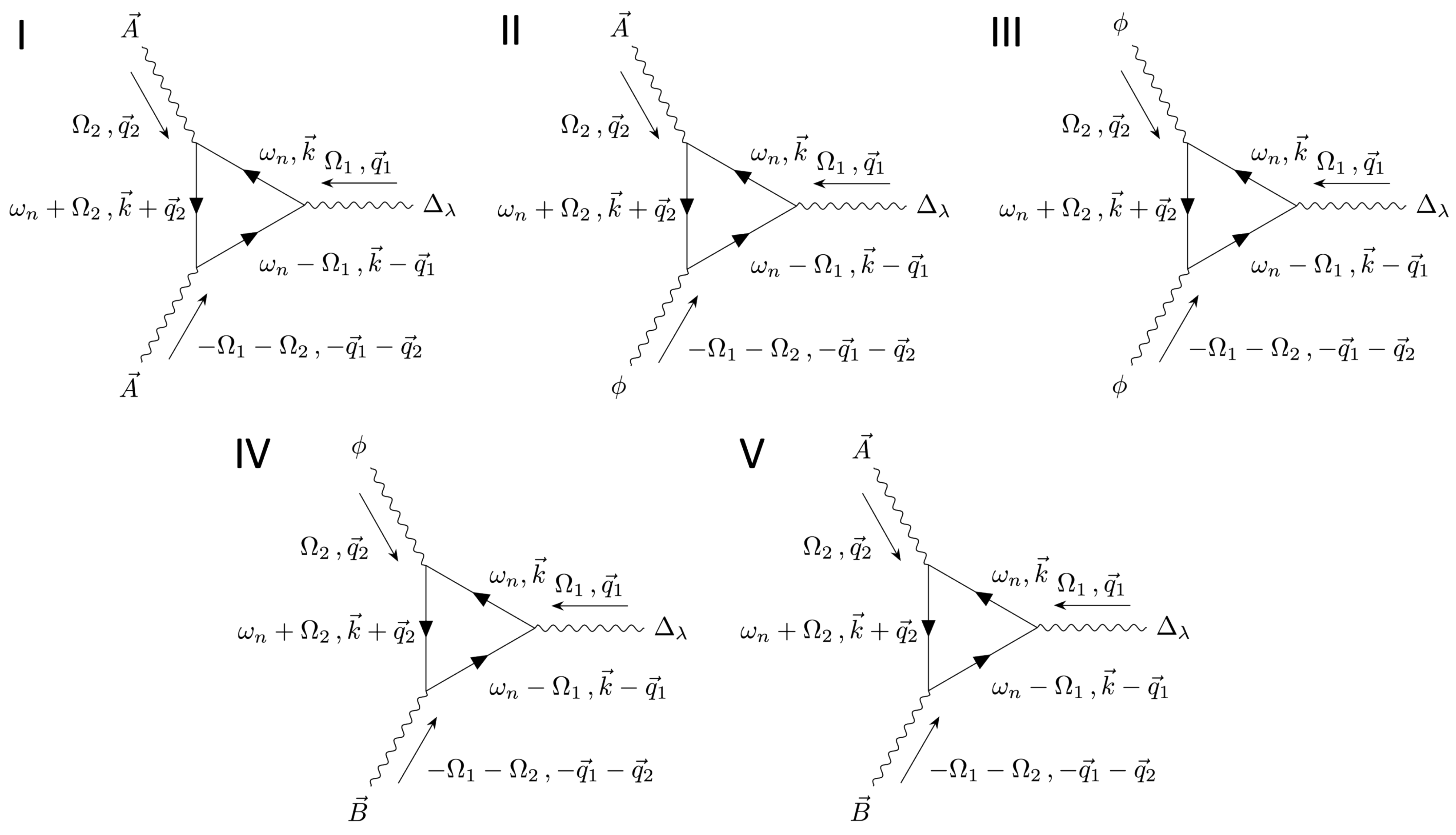}
\caption{Diagrams potentially contributing to axion electrodynamics where $\lambda=s,p$ in $\Delta_\lambda$.
} 
\label{fig:diagrams}
\end{figure*}

\section{Orbital contribution to axion electrodynamics}
\label{sec:orbital}

In this section, we calculate the orbital contribution to the axion action based on the path integral approach.
There are three diagrams which contain two $A^\mu$'s  ($\mu=0,1,2,3$ where $A^0$ is $\phi$) and one $\delta \Delta_\lambda$ ($\lambda=s,p$), as shown in Fig. \ref{fig:diagrams} (I, II, III).
Diagram III -- though not zero -- does not contribute to the axion action, hence we neglect. 
We will only calculate the terms involving two spacetime derivatives in diagrams I, II in Fig. \ref{fig:diagrams}.

\subsection{Diagram I}
\label{subsec:Adot_zero}

This diagram potentially can contribute to $\int d^4 x\partial_t\vec{A}^\prime\cdot \vec{B}$ in the axion action.
However, we demonstrate that in fact, this contribution vanishes.

\subsubsection{$\lambda=s$}

One term contributing to the $\lambda=s$ case is 
$-\frac{1}{6}\text{Tr}[\mathcal{G}_0\Delta H_A^{(1)}\mathcal{G}_0\Delta H_A^{(1)}\mathcal{G}_0\Delta H_s]$.
Including the combinatoric factor of three, we obtain
\begin{flalign}
&D_s^{\text{I}}=\frac{1}{2}\frac{1}{\beta}\sum_{\omega_n}\int \frac{d^3\vec{k}}{(2\pi)^3}
\text{tr}\big[\nn\\
& \mathcal{G}_0(\omega_n,\vec{k})\frac{\hbar}{2m}(2\vec{k}+\vec{q}_2)\cdot \vec{A}^\prime(\Omega_2,\vec{q}_2)
\mathcal{G}_0(\omega_n+\Omega_2,\vec{k}+\vec{q}_2)\nn\\
&\times\frac{\hbar}{2m}(2\vec{k}-\vec{q}_1)\cdot \vec{A}^\prime(-\Omega_1-\Omega_2,-\vec{q}_1-\vec{q}_2)\nn\\
&\times\mathcal{G}_0(\omega_n-\Omega_1,\vec{k}-\vec{q}_1)
\delta \Delta_s(\Omega_1,\vec{q}_1)\gamma^4
\big],
\label{eq:DI_s}
\end{flalign}
in which the minus sign coming from the fermion loop cancels with the sign in Eq. (\ref{eq:S3}), and
\begin{flalign}
\mathcal{G}_0(\omega_n,\vec{k})=\frac{a_\chi(\omega_n,\vec{k})\gamma^\chi}{a_\chi(\omega_n,\vec{k})a_\chi(\omega_n,\vec{k})},
\label{eq:G0}
\end{flalign}
where the summation of $\chi$ is over $\chi=0,1,2,3,4,5$; $\gamma^5=\sigma_0\tau_0$; and
\begin{eqnarray}
a_0=\xi_{\vec{k}},~ a_i=\frac{\Delta_p}{k_f}k_i~(i=1,2,3),~a_4=\Delta_s,~a_5=i\omega_n.
\end{eqnarray}
Plugging Eq. (\ref{eq:G0}) into Eq. (\ref{eq:DI_s}),
the trace of the numerators of $\mathcal{G}_0$ gives  
\begin{flalign}
&\text{tr} \big[
a_{\chi_1}(\omega_n,\vec{k})\gamma^{\chi_1} a_{\chi_2}(\omega_n+\Omega_2,\vec{k}+\vec{q}_2)\gamma^{\chi_2}\nn\\
& a_{\chi_3}(\omega_n-\Omega_1,\vec{k}-\vec{q}_1)\gamma^{\chi_3} \gamma^4
\big]
\label{eq:DI_s_numerator}
\end{flalign}
We note that the $O(\Omega_j)$ ($j=1,2$) term in Eq. (\ref{eq:DI_s_numerator}) is
$8\omega_n\Delta_s(\Omega_1-\Omega_2)$.

To generate $\int d^4 x\partial_t\vec{A}\cdot \nabla \times \vec{A}$, we must consider terms in Eq. (\ref{eq:DI_s}) which involve one $\Omega_j$  and one $\vec{q}_j$ ($j=1,2$).
It is straightforward to see that the $O(\Omega_j)$ term in $1/\sum_{i=0}^5 [a_i(\omega_n+\Omega,\vec{k}+\vec{q})]^2$ ($\Omega=-\Omega_1,\Omega_2$ and $\vec{q}=-\vec{q}_1,\vec{q}_2$) is proportional to $\omega_n\Omega$, similar as the $O(\Omega_j)$ term in the trace in Eq. (\ref{eq:DI_s_numerator}).
Since the Matsubara summation of terms involving odd powers of $\omega_n$ vanishes,
we conclude that there is no contribution to $\int d^4 x\partial_t\vec{A}\cdot \nabla \times \vec{A}$ from Diagram I for $\lambda=s$.

\subsubsection{$\lambda=p$}

The analysis for $\lambda=p$ in Diagram I is exactly similarly
and the contribution again vanishes.

\subsection{Diagram II}

This diagram potentially can contribute to $\int d^4 x\nabla \phi^\prime\cdot \vec{B}$ in the axion action.
We show that it gives exactly the axion action derived in Sec. \ref{subsec:real_space}.

\subsubsection{$\lambda=s$}

One term contributing to the $\lambda=s$ case is 
$-\frac{1}{6}\text{Tr}[\mathcal{G}_0\Delta H_A^{(1)}\mathcal{G}_0\Delta H_\phi\mathcal{G}_0\Delta H_s]$.
Including the combinatoric factor of $6$ and using Eq. (\ref{eq:G0}), we obtain
\begin{flalign}
&D_s^{\text{II}}=\frac{1}{\beta}\sum_{\omega_n}\int \frac{d^3\vec{k}}{(2\pi)^3}\frac{\hbar}{2m}(2\vec{k}+\vec{q}_2)\cdot \vec{A}^\prime(\Omega_2,\vec{q}_2)\nn\\
&\times\phi^\prime(-\Omega_1-\Omega_2,-\vec{q}_1-\vec{q}_2)\delta \Delta_s(\Omega_1,\vec{q}_1)\nn\\
&\times\text{tr} \big[
a_{\chi_1}(\omega_n,\vec{k})\gamma^{\chi_1} a_{\chi_2}(\omega_n+\Omega_2,\vec{k}+\vec{q}_2)\gamma^{\chi_2}\gamma^0\nn\\
&~~~~~~~\cdot a_{\chi_3}(\omega_n-\Omega_1,\vec{k}-\vec{q}_1)\gamma^{\chi_3} \gamma^4
\big],\nn\\
&\times\Pi_{\nu=1}^3\frac{1}{a_\chi(\omega_n+\Omega^\prime_\nu,\vec{k}+\vec{q}^\prime_\nu)a_\chi(\omega_n+\Omega^\prime_\nu,\vec{k}+\vec{q}^\prime_\nu)},
\label{eq:DII_s}
\end{flalign}
in which $\Omega^\prime_1=0$, $\Omega^\prime_2=\Omega_2$, $\Omega^\prime_3=-\Omega_1$,
and $\vec{q}^\prime_1=0$, $\vec{q}^\prime_2=\vec{q}_2$, $\vec{q}^\prime_3=-\vec{q}_1$.
Recall that we want a term $\sim \epsilon_{ijk} \partial_i\phi\partial_jA_k$ in the action.
Such term can be generated from the trace of a multiplication of the product of five $\gamma$-matrices, which is given by
\begin{eqnarray}
\text{tr}(\gamma^0\gamma^i\gamma^j\gamma^k\gamma^4)=-4\epsilon_{ijk}.
\label{eq:tr_epsilon_ijk_2}
\end{eqnarray}
Then the $\epsilon_{ijk}$ term within the trace  in Eq. (\ref{eq:DII_s}) can be evaluated as 
\begin{eqnarray}
\text{tr}[...]&=&-4(\frac{\Delta_p}{k_f})^3\epsilon_{ijk}k_i(k_j+q_{2j})(k_k-q_{1k})\nn\\
&=& 4(\frac{\Delta_p}{k_f})^3\epsilon_{ijk}k_iq_{2j}q_{1k}.
\label{eq:tr_DII_s}
\end{eqnarray}
Since Eq. (\ref{eq:tr_DII_s}) already contains a product of two wavevector $q$'s,
we can set $\vec{q}_j,\Omega_j$ to be zero in the remaining parts of Eq. (\ref{eq:DII_s}).
This gives
\begin{flalign}
&D_s^{\text{II}}=D(\Delta_s,\Delta_p)\epsilon_{ijk}q_{2j}q_{1k}\times\nn\\
&A^\prime_i(\Omega_2,\vec{q}_2)\delta \Delta_s(\omega_1,\vec{q}_1)\phi^\prime(-\Omega_1-\Omega_2,\vec{q}_1-\vec{q}_2),
\label{eq:D_s_1}
\end{flalign}
in which $D(\Delta_s,\Delta_p)$  is given exactly by Eq. (\ref{eq:value_D_2}).
The corresponding term in the action is 
\begin{eqnarray}
\int d\tau d^3x D(\Delta_s,\Delta_p) \phi^\prime\nabla\delta \Delta_s\cdot  \nabla \times \vec{A}^\prime.
\label{eq:axion_II_s}
\end{eqnarray}


\subsubsection{$\lambda=p$}

When $\lambda=p$, $D_p^{\text{II}}$ can be obtained from $D_s^{\text{II}}$ by replacing $\delta\Delta_s$ in Eq. (\ref{eq:DII_s}) with $\delta\Delta_p$, and changing the trace to 
\begin{flalign}
&\text{tr} \big[
a_{\chi_1}(\omega_n,\vec{k})\gamma^{\chi_1} a_{\chi_2}(\omega_n+\Omega_2,\vec{k}+\vec{q}_2)\gamma^{\chi_2}\gamma^0\nn\\
&\cdot a_{\chi_3}(\omega_n-\Omega_1,\vec{k}-\vec{q}_1)\gamma^{\chi_3}  
\frac{1}{k_f} (\vec{k}-\frac{\vec{q}_1}{2})\cdot \vec{\gamma}
\big].
\label{eq:tr_DII_p}
\end{flalign}
The $\epsilon_{ijk}$ term in Eq. (\ref{eq:tr_DII_p})  can be straightforwardly evaluated as 
\begin{eqnarray}
-4\frac{\Delta_s(\Delta_p)^2}{k_f^3}\epsilon_{ijk}q_{2i}q_{1j}k_k.
\end{eqnarray}
As a result, the corresponding term in the axion action in real space is
\begin{eqnarray}
-D^\prime(\Delta_s,\Delta_p) \phi^\prime \nabla \delta\Delta_p \cdot \nabla\times  \vec{A}^\prime,
\label{eq:axion_II_p}
\end{eqnarray}
in which 
\begin{eqnarray}
D^\prime(\Delta_s,\Delta_p)=\frac{\Delta_s}{\Delta_p}D(\Delta_s,\Delta_p).
\end{eqnarray}

\subsection{Diagram III}

This diagram does not contribute to the axion action as explained at the beginning of this section,
though it does contribute to non-axion terms as discussed in Appendix \ref{app:nonaxion}.

\subsection{Orbital contribution to the axion action}

Combining Eq. (\ref{eq:axion_II_s},\ref{eq:axion_II_p}) together and using $\nabla\delta\Delta_\lambda=\nabla\Delta_\lambda$, we obtain
\begin{eqnarray}
S^{o}_{\text{ax}}=\int d\tau d^3x  \phi^\prime(D\nabla\Delta_s-D^\prime\nabla\Delta_p)\cdot \nabla \times \vec{A}^\prime,
\end{eqnarray}
in which $D$ ($D^\prime$) is $D(\Delta_s,\Delta_p)$ ($D(\Delta_s,\Delta_p)$) for short.
Plugging in the expression of $D$ given in Eq. (\ref{eq:value_D_2}), we have
\begin{flalign}
&D\nabla\Delta_s-D^\prime\nabla\Delta_p
=\frac{1}{6\pi^2\hbar} \frac{1}{[1+(\frac{\Delta_s}{\Delta_p})^2]^2}\nabla(\frac{\Delta_s}{\Delta_p}).
\end{flalign}
Using the integral
\begin{flalign}
\int_{\frac{\Delta_s}{\Delta_p}}^\infty \frac{dx}{(1+x^2)^2}=\frac{1}{4}[\pi-2\arctan(\frac{\Delta_s}{\Delta_p})-2\frac{\Delta_s/\Delta_p}{1+(\Delta_s/\Delta_p)^2}],
\label{eq:int_x_orbital}
\end{flalign}
$S^{o}_{\text{ax}}$ becomes
\begin{eqnarray}
S^{o}_{\text{ax}}= -\frac{\alpha}{24\pi^2}\int d\tau d^3x \frac{\phi^\prime}{e} \nabla\Theta^o_{\text{ax}}\cdot \nabla\times\frac{c\vec{A}^\prime}{e},
\label{eq:axion_orbital_orig}
\end{eqnarray}
where $\Theta^o_{\text{ax}}$ coincides exactly with the expression in Eq. (\ref{eq:Theta_ax_expression}).
Integrating by parts, employing Eq. (\ref{eq:prime_A}), and transforming to the real time,
we obtain
\begin{flalign}
S^{o}_{\text{ax}}=\frac{\alpha}{24\pi^2}\int d^4x \Theta^o_{\text{ax}}\nabla(\phi+\frac{\hbar}{e}\partial_t \Phi)\cdot \vec{B},
\label{eq:axion_path_integral}
\end{flalign}
where 
\begin{eqnarray}
\nabla\times\nabla\Phi=0,~\nabla\times\nabla\Theta^o_{\text{ax}}=0, 
\label{eq:vortex_free}
\end{eqnarray}
are used,
which apply to the vortex-free case. 

\section{Zeeman contribution to axion electrodynamics}
\label{sec:Zeeman}

In this section, we calculate the Zeeman contribution to the axion action based on the path integral approach.

\subsection{Diagram IV}

This diagram potentially can contribute to $\int d^4 x\nabla \phi^\prime\cdot \vec{B}$ in the axion action.
We will derive its explicit expression. 

\subsubsection{$\lambda=s$}

One term contributing to the $\lambda=s$ case is 
$-\frac{1}{6}\text{Tr}[\mathcal{G}_0\Delta H_Z\mathcal{G}_0\Delta H_\phi\mathcal{G}_0\Delta H_s]$.
Including the combinatoric factor of $6$, we obtain
\begin{flalign}
&D_s^{\text{IV}}=\frac{1}{\beta}\sum_{\omega_n}\int \frac{d^3\vec{k}}{(2\pi)^3}
\text{tr}\big[\nn\\
& \mathcal{G}_0(\omega_n,\vec{k})\phi^\prime(\Omega_2,\vec{q}_2)\gamma^0
\mathcal{G}_0(\omega_n+\Omega_2,\vec{k}+\vec{q}_2)\nn\\
&\times(-) \frac{1}{4}ig\mu_B\epsilon_{ijk}\gamma^j\gamma^kB_i(-\Omega_1-\Omega_2,-\vec{q}_1-\vec{q}_2)
\nn\\
&\times\mathcal{G}_0(\omega_n-\Omega_1,\vec{k}-\vec{q}_1)
\delta \Delta_s(\Omega_1,\vec{q}_1)\gamma^4
\big].
\label{eq:DI_p}
\end{flalign}
Only considering the axion term $\int d^4 x\nabla \phi^\prime\cdot \vec{B}$, we can set $\Omega_1=\Omega_2=0$, and it is enough to expand $D_s^{\text{IV}}$ up to linear order in $\vec{q}_\alpha$ ($\alpha=1,2$).
Calculations show that ($\alpha,\beta=1,2$; $i,j,k=x,y,z$; $k_i\neq k_j\neq k_k$)
\begin{eqnarray}
D_s^{\text{IV}}=E_{1si}iq_{1i}B_i+E_{2si}iq_{2i}B_i+O(q_{\alpha j}q_{\beta k}),
\end{eqnarray}
in which 
\begin{eqnarray}
E_{\alpha si}&=&-2g\mu_B\frac{\Delta_p}{k_f}\frac{1}{\beta}\sum_{\omega_n}\int \frac{d^3\vec{k}}{(2\pi)^3}M_{\alpha si},
\end{eqnarray}
where
\begin{eqnarray}
M_{1si}&=&\frac{-\omega_n^2+\xi_{\vec{k}}^2+(k_i^2-k_j^2-k_k^2)\frac{\Delta_p^2}{k_f^2}-\Delta_s^2}{[\omega_n^2+\xi_{\vec{k}}^2+(\frac{\Delta_p}{k_f})^2k^2+\Delta_s^2]^3},\nn\\
M_{2si}&=&\frac{
\omega_n^2-\frac{\hbar^2}{2m}(k_f^2-k^2+4k_i^2)\xi_{\vec{k}}+\frac{\Delta_p^2}{k_f^2}(k^2-2k_i^2)-\Delta_s^2
}{[\omega_n^2+\xi_{\vec{k}}^2+(\frac{\Delta_p}{k_f})^2k^2+\Delta_s^2]^3}.\nn\\
\end{eqnarray}
We note that: the $q_{1i}B_j$ ($i\neq j$) terms vanish in $D_s^{\text{IV}}$ after the integration over the solid angle of $\vec{k}$;
and $k_i^2$ can be replaced by $k^2/3$ in the integration.
Therefore, 
\begin{eqnarray}
M_{\alpha s i}\equiv M_{\alpha s},~E_{\alpha s i}\equiv E_{\alpha s},
\end{eqnarray}
where $i=x,y,z$.

We will only keep the leading order terms in an expansion over $\Delta_\lambda/\epsilon_f$ ($\lambda=s,p$).  
Then  in $E_{1s}$, all $k^2$ can be replaced by $k_f^2$ in the integrand $M_{1s}$,
and $\int d^3\vec{k}/(2\pi)^3$ can be set as $N_f \int d\xi$,
where $N_f=\frac{mk_f}{2\pi^2\hbar^2}$. 
On the other hand, $M_{2si}$ contains a term $-\frac{\hbar^2}{2m}(k_f^2+\frac{k^2}{3})\xi_{\vec{k}}$, which is one order less than the other terms in the numerator. 
Therefore, in this case, $\int d^3\vec{k}/(2\pi)^3$ should be set as $N_f \int (1+\frac{\xi}{2\epsilon_f}) d\xi$,  and 
\begin{flalign}
&M_{2s}\rightarrow 
-\frac{\frac{4}{3}\xi_{\vec{k}}\epsilon_f}{[\omega_n^2+\xi_{\vec{k}}^2+(\frac{\Delta_p}{k_f})^2k^2+\Delta_s^2]^3}+\nn\\
&\frac{
\omega_n^2-\frac{1}{3}\xi_{\vec{k}}^2+\frac{1}{3}\Delta_p^2-\Delta_s^2
}{[\omega_n^2+\xi_{\vec{k}}^2+(\frac{\Delta_p}{k_f})^2k^2+\Delta_s^2]^3}
+\frac{4\Delta_p^2\xi_{\vec{k}}^2}{[\omega_n^2+\xi_{\vec{k}}^2+(\frac{\Delta_p}{k_f})^2k^2+\Delta_s^2]^4},
\label{eq:M2s_2}
\end{flalign}
where the first term can combine with $N_f \int \frac{\xi}{2\epsilon_f} d\xi$ within $\int d^3\vec{k}/(2\pi)^3$ giving a nonzero contribution, 
and the third term in Eq. (\ref{eq:M2s_2}) comes from expanding the denominator using $\Delta_p^2k^2/k_f^2=\Delta_p^2(1+\xi_{\vec{k}}/\epsilon_f)$.
Again at zero temperature, $\frac{1}{\beta}\sum_{\omega_n}=\int \frac{d\omega}{2\pi}$.
Performing the integrations $\int d\omega\int d\xi$, we obtain
\begin{eqnarray}
E_{1s}(\Delta_s,\Delta_p)&=&\frac{g\mu_Bm}{12\pi^2\hbar^2}\frac{\Delta_p(3\Delta_s^2+\Delta_p^2)}{(\Delta_p^2+\Delta_s^2)^2},\nn\\
E_{2s}(\Delta_s,\Delta_p)&=&\frac{g\mu_Bm}{4\pi^2\hbar^2}\frac{\Delta_p(\Delta_s^2-\Delta_p^2)}{(\Delta_p^2+\Delta_s^2)^2}.
\end{eqnarray}

Transforming back to the real space, the action becomes
\begin{flalign}
S_{\text{ax}}^Z=\int d\tau d^3x\big[ E_{1s} \phi^\prime \nabla \delta \Delta_s\cdot \vec{B} +E_{2s} \delta \Delta_s \vec{B}\cdot\nabla \phi^\prime
\big].
\end{flalign}
Integrating by parts,
we obtain the axion action
\begin{eqnarray}
&\int d\tau d^3x (E_{1s}-E_{2s}) \phi^\prime\nabla \delta \Delta_s\cdot  \vec{B}\nn\\
&+\int d\tau d^3x E_{2s} \vec{B}\cdot \nabla (\phi^\prime  \delta \Delta_s).
\label{eq:axion_spin_s}
\end{eqnarray}

\subsubsection{$\lambda=p$}

Similar as the $\lambda=s$ case, the expression with a spatially varying  $\delta \Delta_p$ is 
\begin{flalign}
&D_p^{\text{IV}}=\frac{1}{\beta}\sum_{\omega_n}\int \frac{d^3\vec{k}}{(2\pi)^3}
\text{tr}\big[\nn\\
& \mathcal{G}_0(\omega_n,\vec{k})\phi^\prime(\Omega_2,\vec{q}_2)\gamma^0
\mathcal{G}_0(\omega_n+\Omega_2,\vec{k}+\vec{q}_2)\nn\\
&\times(-) \frac{1}{4}ig\mu_B\epsilon_{ijk}\gamma^j\gamma^kB_i(-\Omega_1-\Omega_2,-\vec{q}_1-\vec{q}_2)
\nn\\
&\times\mathcal{G}_0(\omega_n-\Omega_1,\vec{k}-\vec{q}_1)
\frac{1}{k_f}\delta \Delta_p(\Omega_1,\vec{q}_1)(\vec{k}-\frac{\vec{q}_1}{2})\cdot \vec{\gamma} 
\big].
\label{eq:D4_p}
\end{flalign}
Again setting $\Omega_1=\Omega_2=0$, and 
keeping the linear in $\vec{q}_\alpha$ ($\alpha=1,2$), we obtain
\begin{eqnarray}
D_p^{\text{IV}}=E_{1p}iq_{1i}B_i+E_{2p}iq_{2i}B_i+O(q_{\alpha j}q_{\beta k}),
\end{eqnarray}
in which 
\begin{eqnarray}
E_{\alpha p}&=&-g\mu_B\frac{\Delta_s}{k_f}\frac{1}{\beta}\sum_{\omega_n}\int \frac{d^3\vec{k}}{(2\pi)^3}M_{\alpha p},
\end{eqnarray}
where ($k_i\neq k_j\neq k_k$)
\begin{eqnarray}
M_{1p}&=&\frac{\omega_n^2+\xi_{\vec{k}}^2+(\frac{\Delta_p}{k_f})^2(k^2-4k_i^2)+\Delta_s^2 }{[\omega_n^2+\xi_{\vec{k}}^2+(\frac{\Delta_p}{k_f})^2k^2+\Delta_s^2]^3},\nn\\
M_{2p}&=&\frac{4[ -(\frac{\Delta_p}{k_f})^2(k_j^2+k_k^2) +\frac{\hbar^2}{m}k_i^2\xi_{\vec{k}}]
}{[\omega_n^2+\xi_{\vec{k}}^2+(\frac{\Delta_p}{k_f})^2k^2+\Delta_s^2]^3}.
\label{eq:Mp}
\end{eqnarray}
We note that again, the values of Eq. (\ref{eq:Mp}) do not depend on $i=x,y,z$. 
Calculations show that (up to leading order in $\Delta_\lambda/\epsilon_f$, where $\lambda=s,p$)
\begin{eqnarray}
E_{\alpha p}=-\frac{\Delta_s}{\Delta_p}E_{\alpha s}.
\label{eq:Ep_spin}
\end{eqnarray}
Correspondingly, the contribution to the axion action is 
\begin{eqnarray}
&\int d\tau d^3x (E_{1p}-E_{2p}) \phi^\prime\nabla \delta \Delta_p\cdot\vec{B} \nn\\
&+\int d\tau d^3x E_{2p} \vec{B}\cdot \nabla (\phi^\prime  \delta \Delta_p).
\label{eq:axion_spin_p}
\end{eqnarray}

\subsection{Diagram V}

This diagram potentially can contribute to $\int d^4 x\partial_t\vec{A}^\prime\cdot \vec{B}$ in the axion action.
However, in a way similar with the discussions in Sec. \ref{subsec:Adot_zero},
it can be seen that the axion contribution from this diagram vanishes, since the integrand is odd with respect to the fermionic Matsubara frequency.

\subsection{Zeeman contribution to the axion action}

Combining Eqs. (\ref{eq:axion_spin_s}, \ref{eq:axion_spin_p}) and using Eq. (\ref{eq:Ep_spin}),
we obtain
\begin{eqnarray}
S_{\text{ax}}^Z&=&\int d\tau d^3x  \phi^\prime E^Z(\Delta_s,\Delta_p)\nabla(\frac{\Delta_s}{\Delta_p})\cdot \vec{B}\nn\\
&&+\int d\tau d^3x E_2^Z (\Delta_s,\Delta_p)  \vec{B}\cdot \nabla \big[\phi^\prime  \delta (\frac{\Delta_s}{\Delta_p})\big],
\label{eq:S_ax_Z_0}
\end{eqnarray}
in which 
\begin{eqnarray}
E^Z&=&\Delta_p(E_{1s}-E_{2s})\nn\\
&=&\frac{g\mu_B m}{3\pi^2\hbar^2}\frac{1}{[1+(\Delta_s/\Delta_p)^2]^2},
\end{eqnarray}
and
\begin{eqnarray}
E_2^Z&=&\Delta_p E_{2s}\nn\\
&=&\frac{g\mu_B m}{4\pi^2\hbar^2}\frac{(\Delta_s/\Delta_p)^2-1}{[1+(\Delta_s/\Delta_p)^2]^2}.
\end{eqnarray}
Since 
\begin{eqnarray}
E_2^Z (\Delta_s,\Delta_p)\delta \big(\frac{\Delta_s}{\Delta_p}\big)=\frac{g\mu_B m}{4\pi^2\hbar^2}\delta \big(\frac{\Delta_s/\Delta_p}{1+(\Delta_s/\Delta_p)^2}\big)
\end{eqnarray}
and $\nabla\cdot \vec{B}=0$, the second term in Eq. (\ref{eq:S_ax_Z_0}) can be converted to an integral of a total derivative, which vanishes. 
Therefore, it is enough to only keep the first term in Eq. (\ref{eq:S_ax_Z_0}).

Again imposing the condition $\Theta_{\text{ax}}=0$ for the pure $s$-wave case
and using 
\begin{flalign}
&\int_{\frac{\Delta_s}{\Delta_p}}^{+\infty} \frac{dx}{(1+x^2)^2}=\\
&\frac{1}{2}\big[\frac{\pi}{2}-\arctan(\Delta_s/\Delta_p)-\frac{\Delta_s/\Delta_p}{1+(\Delta_s/\Delta_p)^2}\big],
\label{eq:int_x_spin}
\end{flalign}
$S_{\text{ax}}^Z$ becomes 
\begin{eqnarray}
S_{\text{ax}}^Z=-\frac{e}{24\pi^2\hbar c}\int d\tau d^3x \phi^\prime\nabla \Theta^Z_{\text{ax}} \cdot \vec{B},
\label{eq:axion_zeeman_orig_2}
\end{eqnarray}
in which $\mu_B$ is replaced by $\frac{e\hbar}{2mc}$, $\alpha $ is the fine structure constant,  and
\begin{eqnarray}
\Theta^Z_{\text{ax}}=g(\pi-2\arctan (\frac{\Delta_s}{\Delta_p})-\frac{2\Delta_s\Delta_p}{\Delta_s^2+\Delta_p^2}).
\end{eqnarray}
Further performing an integration by part and transforming to real time, we obtain
\begin{eqnarray}
S_{\text{ax}}^Z=\frac{\alpha}{24\pi^2}\int d^4x \Theta^Z_{\text{ax}} \nabla (\phi+\frac{\hbar}{e}\partial_t \Phi)\cdot \vec{B}.
\label{eq:S_ax_Z}
\end{eqnarray}
in which  the Land\'e factor $g$ is equal to $2$ in vacuum, 
but can be somewhat arbitrary in  solid state materials. 
The action $S_{\text{ax}}^Z$ in Eq. (\ref{eq:S_ax_Z}) can also contribute to transverse supercurrent, similar to its orbital counterpart as discussed in Sec. \ref{subsec:real_space}.

\section{Physical effects}
\label{sec:physical}

Including both the orbital and Zeeman contributions in Eq. (\ref{eq:axion_path_integral}) and Eq. (\ref{eq:S_ax_Z}), we obtain  the axion action in Eq. (\ref{eq:Full_axion_action_1}).
To obtain the full effective action $S_{\text{eff}}$ to lowest orders, we need to include the quadratic terms for the phase mode $\Phi$ \cite{Altland2010}, which gives
\begin{align}
    S_{\text{eff}} =& \int d^4x [\frac{N_f}{2}(\hbar\partial_t\Phi+e\phi)^2 - \frac{n_s}{2m}(\hbar\nabla\Phi+\frac{e \vec{A}}{c})^2]\nn\\
    &+\frac{\alpha}{24\pi^2}\int d^4x ~\Theta_{\text{ax}}~ \nabla(\phi+\frac{\hbar\partial_t\Phi}{e})\cdot \vec{B},
    \label{eq:Seff}
\end{align}
in which $N_f$ is the density of states at Fermi energy, $n_s$ is the superfluidity density (equal to electron density at zero temperature),
and $\Theta_{\text{ax}}=\Theta^o_{\text{ax}}+\Theta^Z_{\text{ax}}$ defined in Eq. (\ref{eq:def_Theta}).  
We note that the axion field is massive, hence the mass term of $\Theta_{\text{ax}}$ is neglected in the low energy effective action. 

In this section, we will discuss the physical effects derived from the effective action,
including the current inflow to the vortex line and the Witten effect.
We emphasize that different from usual superconductors, a vortex line here refers to a vortex line of $\Theta$, not  of the superconducting phase $\Phi$.

\subsection{Current inflow to the vortex  on the superconducting surface}

Recall that in the derivation of Eq. (\ref{eq:Seff}), we have used the vortex-free condition for both $\Phi$ and $\Theta$ to perform integration by parts.
In the presence of vortices, the expression before integration by parts should be used (see Eq. (\ref{eq:axion_orbital_orig}) and Eq. (\ref{eq:axion_zeeman_orig_2})), which gives
\begin{align}
S^\prime_{\text{eff}} &= \int d^4x [\frac{N_f}{2}(\hbar\partial_t\Phi+e\phi)^2 - \frac{n_s}{2m}(\hbar\nabla\Phi+\frac{e \vec{A}}{c})^2]\nn\\
&-\frac{\alpha}{24\pi^2}\int d^4x (\phi+\frac{\hbar\partial_t\Phi}{e}) \nabla\Theta_{\text{ax}}\cdot \vec{B}.
\label{eq:Seff_prime}
\end{align}

The electric charge  can be obtained from $S^\prime_{\text{eff}}$ as
\begin{eqnarray}
\rho& = &\frac{\delta S^\prime_{\text{eff}}}{\delta \phi}\nn\\
&=&e N_f (\hbar\partial_t \Phi + e\phi) -\frac{\alpha}{24\pi^2}\nabla\cdot(\Theta_{\text{ax}} \vec{B})
\label{eq:nres}
\end{eqnarray}
where $\nabla\cdot \vec{B}=0$ and $\nabla\times\nabla\Phi=0$ (since we assume that there is no vortex of $\Phi$) are used.
Similarly, the electric current is
\begin{eqnarray}
\vec{j} = c\frac{\delta S^\prime_{\text{eff}}}{\delta \vec{A}} =\vec{j}^{(1)}+\vec{j}^{(2)}+\vec{j}^{(3)},
\label{eq:response}
\end{eqnarray}
in which 
\begin{eqnarray}
\vec{j}^{(1)}&=& -\frac{en_s}{m}(\frac{e\vec{A}}{c} -\hbar\nabla \Phi) \nn\\
\vec{j}^{(2)}&=& -\frac{\alpha c}{24\pi^2}\nabla (\phi+\frac{\hbar\partial_t\Phi}{e})\times\nabla\Theta_{\text{ax}}\nn\\
\vec{j}^{(3)}&=&-\frac{\alpha c}{24\pi^2}(\phi+\frac{\hbar\partial_t\Phi}{e})\nabla\times\nabla\Theta_{\text{ax}}.
\label{eq:j123}
\end{eqnarray}

We note that in the absence of any vortex, continuity equation $\partial_t\rho +\nabla\cdot \vec{j}=0$
is ensured by  the equation of motion of $\Phi$
\begin{equation}
\sum_{\mu=t,x,y,z}\partial_\mu \frac{\partial \mathcal{L}}{\partial\partial_\mu \Phi} = 0,
\end{equation}
and the relation  $\nabla \cdot \vec{j}^{(2)}=0$ (which holds when $\nabla\times \nabla \Theta_{\text{ax}}=0$). 
On the other hand, in the presence of a vortex of $\Theta_{\text{ax}}$, 
$\nabla\cdot (\vec{j}^{(2)}+\vec{j}^{(3)})$ is not included in the equation of motion of $\Phi$.
However, this additional term vanishes since 
\begin{flalign}
\nabla\cdot (\vec{j}^{(2)}+\vec{j}^{(3)})=\nabla \cdot \big[\nabla\times \big((\phi+\frac{\hbar\partial_t\Phi}{e})\nabla\Theta_{\text{ax}}\big)\big]=0.
\end{flalign}

\begin{figure}
    \centering
    \includegraphics[width=8cm]{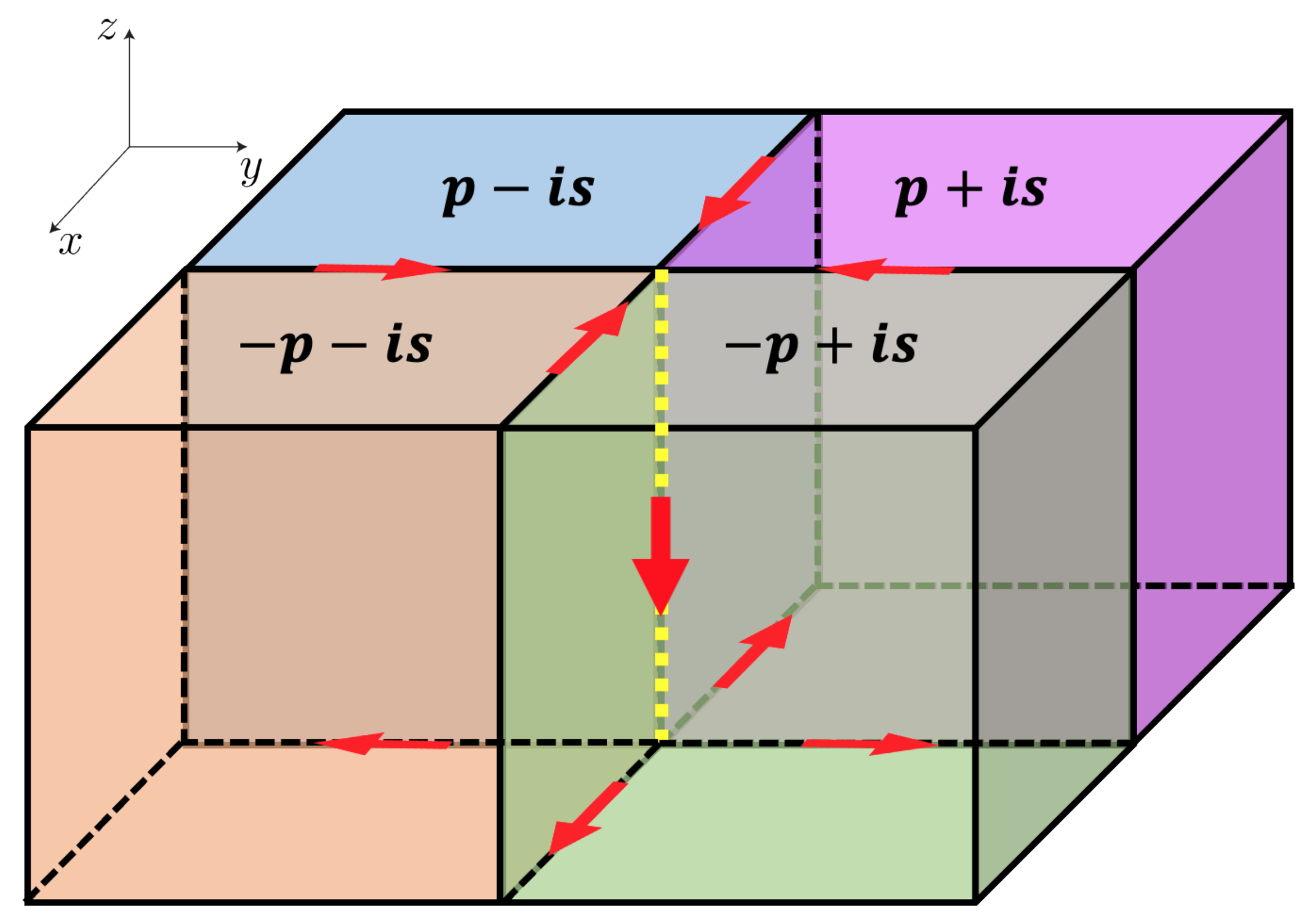}
    \caption{Electric currents on the surfaces of the superconducting bulk and along the vortex line, when a vortex line of $\Theta_{\text{ax}}$ is present. 
    The red arrows denote the directions of the electric currents, and the yellow dashed line represents the location of the vortex core. 
    }
    \label{fig:string}
\end{figure}

To understand the physical meaning of Eq. (\ref{eq:response}), we consider a vortex line of $\Theta_{\text{ax}}$ along $z$-direction  
as shown in Fig. \ref{fig:string},
where the $xy$-coordinates of the vortex core are assumed to be $x=y=0$,
as represented by the yellow dashed line in Fig. \ref{fig:string}. 
Fig. \ref{fig:string} represents a vortex line since the axion angles are different in $p+is$, $p-is$, $-p+is$, $-p-is$ bulks, 
and in fact, $\Theta_{\text{ax}}$ winds by $2\pi$ when these four bulks are successively traversed.

In addition, the surface of the superconducting bulk is the place where a sudden jump of $\phi$ occurs,
since the vacuum can be viewed as having very large electric potential such that all electrons are depleted. 
Hence, both $\nabla\Theta_{\text{ax}}$ and $\nabla \phi$ are nonzero at the upper surface of the superconducting bulk.
By virtue of $\vec{j}^{(2)}$ in Eq. (\ref{eq:j123}), this leads to surface currents flowing to the vortex line as shown by the four red arrows on the upper surface in Fig. \ref{fig:string}.
The total inflowing current into the vortex line can be obtained by performing a line integral of $\vec{j}^{(2)}$ over a loop surrounding the vortex core on the upper surface, which gives
\begin{equation}
\dot{N} = \frac{\alpha c}{12\pi }\partial_z(\phi+\frac{\hbar\partial_t\Phi}{e}),
\label{eq:dot_N_1}
\end{equation}
where $\dot{N}$ represents the particle number injected into the vortex line per unit time. 

Charge conservation is recovered  $\vec{j}^{(3)}$ in Eq. (\ref{eq:j123}).
Notice that $\nabla\times\nabla\Theta$ is proportional to $\delta(x)\delta(y)$, hence it only has non-vanishing effects on the vortex line. 
Since $\nabla\times\nabla\Theta$ is parallel to the $z$-direction, an integration $\int dx dy \vec{j}^{(3)}$ gives the total current running along the vortex line.
Performing the integral, we obtain
\begin{align}
  \int dx dy  \vec{j}^{(3)} &= -\frac{\alpha c}{12\pi }(\phi+\frac{\hbar\partial_t\Phi}{e})\hat{z}.
  \label{eq:int_j3}
\end{align}
The divergence of Eq. (\ref{eq:int_j3}) gives the rate of change of particle number $\dot{N}$ as
\begin{align}
\dot{N} = \nabla\cdot \big(  \int dx dy  \vec{j}^{(3)}\big)= -\frac{\alpha c}{12\pi }\partial_z(\phi+\frac{\hbar\partial_t\Phi}{e}),
\label{eq:dot_N_2}
\end{align}
which exactly cancels with Eq. (\ref{eq:dot_N_1}),
ensuring the condition of charge conservation. 
Notice that $\dot{N}$ in Eq. (\ref{eq:dot_N_2}) is nonzero only close to the upper surface. 
This is physically correct:
The downward flowing current along the vortex line quickly increases as we go down the vortex line from the surface, since there are inflowing currents due to Eq. (\ref{eq:dot_N_1});
on the other hand, deep in the bulk, the current reaches a steady flow along the vortex line, giving rise to a divergence free condition $\dot{N}=0$.

\subsection{Witten effect}

The second term in the density response (Eq.~\eqref{eq:nres}) indicates a  Witten effect in this system. 
Assuming  a magnetic monopole at the origin of the bulk,  we have $\nabla\cdot \vec{B}\propto \delta^{(3)}(\vec{r})$. 
An adiabatic change of a homogenous $\Theta_{\text{ax}}$ by $\Delta\Theta_{\text{ax}}$ leads to an accumulation of particles $\Delta Q$ at the origin, 
which can be derived as
\begin{equation}
    \Delta Q = \int d^3x dt \,\partial_t\rho = \int d^3xdt\,\partial_t\Theta_{\text{ax}}\, \nabla\cdot \vec{B}\propto \Delta\Theta_{\text{ax}}.
\end{equation}

\section{Conclusion}
\label{sec:conclusion}

In conclusion, we have studied the coupling between the axion field and the electromagnetic field in $p+is$ superconductors. 
We find that the axion electrodynamics in $p+is$ superconductors exhibits a nonrelativistic form,
which is different from the superconducting Dirac or Weyl systems. 
As applications of the derived nonrelativistic axion action, the vortex lines of the axion angle and Witten effect are discussed. 
Our work reveals the crucial differences for axion electrodynamics between the $p+is$ superconductors and the superconducting Dirac/Weyl systems. 

{\it Acknowledgments}
CX is supported by Strategic Priority Research Program
of CAS (No. XDB28000000) and The Office of Naval Research under Grant No. N00014-18-1-2722.
WY is supported by the postdoctoral fellowship at Stewart Blusson Quantum Matter Institute, University of British Columbia.

\appendix

\section{Real space calculation of transverse supercurrent}
\label{sec:transverse_current}

In this appendix, we perform a real space calculation  of the transverse supercurrent induced by static electric field and spatial inhomogeneity of $\Delta_s$.
The corresponding term in the axion action $S_{\text{ax}}$ can then be determined from the relation $j_i=-c\frac{\delta S_{ax}}{\delta A_i}$ ($i=1,2,3$) where $c$ is the light velocity.
In addition to the orbital supercurrent discussed in this section,
there is also bound current originating from the spin magnetic moment, as discussed in detail in Sec. \ref{sec:Zeeman}.

Recall that the expectation value of the current operator is given by Eq. (\ref{eq:j_expectation}),
in which the Green's function $\hat{G}$ can be expressed as Eq. (\ref{eq:G_square}),
where $H^2$ is given by Eq. (\ref{eq:H_2}) and Eq. (\ref{eq:H0_2}).
Straightforward calculations show that
\begin{eqnarray}
\hat{T}^\rho \hat{T}^\rho&=&[-\frac{\hbar^2}{2m}\nabla^2-\mu(\vec{r})]^2+\frac{\Delta_p^2}{k_f^2}(-\nabla^2)+[\Delta_s(\vec{r})]^2,\nn\\
 \text{$[$}\hat{T}^0,\hat{T}^i \text{$]$}&=&-i\frac{\Delta_p}{k_f} \partial_i\mu,\nn\\
 \text{$[$}\hat{T}^0,\hat{T}^4 \text{$]$}&=&-\frac{\hbar^2}{2m}(\nabla^2\Delta_s+2\nabla\Delta_s\cdot \nabla),\nn\\
  \text{$[$}\hat{T}^i,\hat{T}^4 \text{$]$}&=&-i\frac{\Delta_p}{k_f} \partial_i\Delta_s,
\end{eqnarray}
in which $\mu(\vec{r})=\epsilon_F+e\phi(\vec{r})$,
where $\phi(\vec{r})$ is the electric potential.

The $n=0$ term in Eq. (\ref{eq:j_expansion}) vanishes:
The odd-in-$\partial_\tau$ term vanishes after Matsubara frequency summation;
and the other terms contain a trace of a single $\gamma$-matrix, hence are also zero.

The $n=1$ term also vanishes.
Removing the odd-in-$\partial_\tau$ term, the $n=1$ term contains one $\Delta_H$ and one $H$ under the trace operation.
However, $\Delta_H$ and $H$ are products of two and one $\gamma$-matrices, respectively.
Hence the result vanishes since the trace of a product of three $\gamma$-matrices is zero.

Finally we consider the $n=2$ term
\begin{flalign}
&\left<\hat{j}_i^{(2)}(\vec{x})\right>=\nn\\
&-\text{Tr} \big[ \hat{j}_i(\vec{x}) 
\frac{1}{-\partial_\tau^2+H_0^2}Q\frac{1}{-\partial_\tau^2+H_0^2}Q\frac{1}{-\partial_\tau^2+H_0^2}H
 \big].
\end{flalign}
To lowest order in the gradient expansion \cite{Niemi1986}, $\mu(\vec{r})$  and $ \Delta_s(\vec{r})$ can be taken as constants in $H_0$ \cite{Niemi1986}. 
The only way to have a nonzero trace is a multiplication of all the five $\gamma$-matrices.
Since the electric current operator $\hat{j}_i(\vec{x})$ contains a $-i\partial_i$, 
the $H$ term must contribute $\frac{\Delta_p}{k_f}(-i\partial_i)\gamma^i$ so that the trace is nonzero.
Therefore, we obtain 
\begin{eqnarray}
&\left<\hat{j}_i^{(2)}(\vec{x})\right>=2(\frac{\Delta_p}{k_f})^3 \text{Tr}\big[
\hat{j}_i(\vec{x}) \frac{1}{-\partial_\tau^2+H_0^2}\partial_j\mu\frac{1}{-\partial_\tau^2+H_0^2}\partial_k \Delta_s \nn\\
&\times\frac{1}{-\partial_\tau^2+H_0^2}(-i\partial_i)
\big] \text{tr}(\gamma^0\gamma^j\gamma^k\gamma^4\gamma^i).
\label{eq:j_2_a}
\end{eqnarray}
in which the trace of the product of five $\gamma$-matrices can be easily evaluated using
\begin{eqnarray}
\text{tr}(\gamma^0\gamma^i\gamma^j\gamma^k\gamma^4)=-4\epsilon_{ijk}.
\label{eq:tr_epsilon_ijk}
\end{eqnarray}

Using Eq. (\ref{eq:def_j_bdg}), $\left<\hat{j}_i^{(2)}(\vec{x})\right>$ can be evaluated as
\begin{flalign}
&\left<\hat{j}_i^{(2)}(\vec{x})\right>=-\epsilon_{ijk}\frac{2e\hbar}{m}(\frac{\Delta_p}{k_f})^3\int d\vec{y}d\vec{z}\nn\\
&\big\{\bra{\vec{x}}(-i\partial_i)\frac{1}{-\partial_\tau^2+H_0^2}\ket{\vec{y}}\partial_j\mu(\vec{y})\bra{\vec{y}}(-i\partial_i)\frac{1}{-\partial_\tau^2+H_0^2}\ket{\vec{z}}\nn\\
&~~~~~~\times\partial_k\Delta_s(\vec{z})\bra{\vec{z}}\frac{1}{-\partial_\tau^2+H_0^2}(-i\partial_i)\ket{\vec{x}}\nn\\
&+\bra{\vec{x}}\frac{1}{-\partial_\tau^2+H_0^2}\ket{\vec{y}}\partial_j\mu(\vec{y})
\bra{\vec{y}}\frac{1}{-\partial_\tau^2+H_0^2}\ket{\vec{z}}\partial_k\Delta_s(\vec{z})\nn\\
&~~~~~~\times\bra{\vec{z}}\frac{1}{-\partial_\tau^2+H_0^2}(-i\partial_i)^2\ket{\vec{x}}\big\}.
\label{eq:j2_realspace}
\end{flalign}
In momentum space, Eq. (\ref{eq:j2_realspace}) becomes
\begin{flalign}
&\left<\hat{j}_i^{(2)}(\vec{x})\right>=-\epsilon_{ijk}\frac{2e\hbar}{m}(\frac{\Delta_p}{k_f})^3\nn\\
&\times\frac{1}{\beta}\sum_{\omega_n}\int d\vec{y}d\vec{z}\partial_j\mu(y)\partial_k\Delta_s(z)\big[\Pi_{\alpha=1}^3 \int \frac{d\vec{k}_\alpha}{(2\pi)^3} \big]\nn\\
&\times(k_{1i}k_{3i}+k_{3i}^2)\big[\Pi_{\alpha=1}^3
\frac{1}{\omega_n^2+[H_0(\vec{k}_\alpha)]^2} e^{ik_\alpha\cdot(\vec{r}_\alpha-\vec{r}_{\alpha+1})}\big],
\label{eq:j2_realspace_2}
\end{flalign}
in which 
$\beta$ is the inverse temperature;
$\omega_n=(2n+1)\pi/\beta$ is the fermionic Matsubara frequency;
 $\vec{r}_{1,2,3}=\vec{x},\vec{y},\vec{z}$;
and $\vec{r}_{4}=\vec{r}_1$.

To lowest order in the gradient expansion, we can set $\vec{y}=\vec{z}$ in $\partial_j\mu(\vec{y})\partial_k\Delta_s(\vec{z})$ within Eq. (\ref{eq:j2_realspace_2}),
then integrating over $\vec{z}$ gives a momentum delta function $\delta^{(3)}(\vec{k}_2-\vec{k}_3)$.
Furthermore, if the higher order terms in the gradient expansion are neglected,  then $\vec{y}$ can be set as  $\vec{x}$ in $\partial_j\mu(\vec{y})\partial_k\Delta_s(\vec{y})$,
and the integration over $\vec{y}$ gives  $\delta^{(3)}(\vec{k}_1-\vec{k}_2)$.
As a result, we obtain $j_i(\vec{x})=\sum_{n=0}^2\left<\hat{j}_i^{(n)}(\vec{x})\right>$ as
\begin{eqnarray}
j_i(\vec{x})=e^2D(\Delta_s,\Delta_p)\epsilon_{ijk}\partial_j\phi(\vec{x})\partial_k\Delta_s(\vec{x}),
\label{eq:j_i_expression}
\end{eqnarray}
in which $\partial_j \mu=-e\partial_j\phi$ is used, and the coefficient $D$ is
\begin{flalign}
&D(\Delta_s,\Delta_p)=\frac{4\hbar}{m}(\frac{\Delta_p}{k_f})^3\times\nn\\
&\frac{1}{\beta}\int \frac{d^3k}{(2\pi)^3}  \frac{k^2/3}{[\omega_n^2+\xi^2(\vec{k})+(\frac{\Delta_p}{k_f})^2k^2+\Delta_s^2]^3},
\label{eq:Coeff_D}
\end{flalign}
where the replacement $k_i^2\rightarrow k^2/3$ is used which holds under integration.
For simplicity, we consider zero temperature such that $\frac{1}{\beta}\sum_n$ can be replaced by $\int \frac{d\omega}{2\pi}$.
In the weak pairing limit $\Delta_s/\epsilon_F,\Delta_p/\epsilon_F\ll 1$, 
$k$ can be simply taken as $k_f$ in $(\frac{\Delta_p}{k_f}k)^2$ in the denominator,
and $\int dk=\int d\xi/(\hbar v_f)$ where $v_f=\hbar k_f/m$ is the Fermi velocity.
Eq. (\ref{eq:Coeff_D}) can then be evaluated, yielding
\begin{eqnarray}
D(\Delta_s,\Delta_p)=\frac{1}{6\pi^2\hbar} \frac{\Delta_p^3}{(\Delta_p^2+\Delta_s^2)^2},
\label{eq:value_D}
\end{eqnarray}
which is Eq. (\ref{eq:value_D_2}).


\section{Symmetry allowed non-axion terms}
\label{app:nonaxion}

In this appendix we examine 
all the symmetry allowed terms which contain one $\Delta_\lambda$ ($\lambda=s,p$),
two $A^\mu$'s ($\mu=t,x,y,z$),
and two spacetime derivatives. 
Up to an overall factor $F(\Delta_s^2,\Delta_s/\Delta_p)\Delta_\lambda$ (where $F$ is a function which can be determined by calculating the corresponding diagram), 
the terms invariant under 3D rotations and $\mathcal{PT}$-operation are
\begin{align}
    &\partial_t\vec{A}^\prime \cdot \partial_t \vec{A}^\prime ~ \delta\Delta_\lambda\nn\\
    &\partial_t^2\vec{A}^\prime \cdot \vec{A}^\prime ~ \delta\Delta_\lambda\nn\\
    &\partial_t\phi^\prime \partial_t\phi^\prime ~ \delta\Delta_\lambda\nn\\
    &\partial_t^2\phi^\prime ~\phi^\prime ~ \delta\Delta_\lambda
\end{align}
\begin{align}
    &(\nabla\cdot\vec{A}^\prime) (\nabla \cdot\vec{A}^\prime )~ \delta\Delta_\lambda\notag\\
    &\nabla^2\vec{A}^\prime \cdot \vec{A}^\prime ~ \delta\Delta_\lambda\nn\\
    &\nabla\phi^\prime \cdot\nabla\phi^\prime ~ \delta\Delta_\lambda\notag\\
    &\nabla^2\phi^\prime ~ \phi^\prime ~\delta \Delta_\lambda
\end{align}
\begin{align}
    &\partial_t \vec{A}^\prime \cdot(\nabla\times\vec{A}^\prime)~\delta\Delta_\lambda\nn\\
    &\vec{A}^\prime \cdot(\nabla\times\partial_t\vec{A}^\prime)~\delta\Delta_\lambda\label{eq:d1_2}
\end{align}
\begin{flalign}
    &(\nabla\cdot\vec{A}^\prime) \partial_t\phi^\prime~ \delta\Delta_\lambda\label{eq:d2_1}
\end{flalign}
\begin{flalign}
    &(\nabla\times\vec{A}^\prime)\cdot(\nabla\phi^\prime)~\delta\Delta_\lambda\label{eq:d2_ax}\,,
\end{flalign}
in which $\lambda=s,p$.
The terms in Eq.~\eqref{eq:d1_2} vanish as discussed in Sec. \ref{subsec:Adot_zero},
and Eq.~\eqref{eq:d2_ax} is the axion term  which has been  calculated and discussed in the main text.

Here we make a comment on the order of the coefficients of the non-axion terms. 
In superconductors, the leading term in the action for the phase mode is
\begin{eqnarray}
\int d^4x \big[N_f (\hbar\partial_t\Phi+e\phi)^2-\frac{n_s}{2m} (\hbar\nabla \Phi+\frac{e\vec{A}}{ c})^2\big],
\end{eqnarray}
in which $n_s$ is the superfluid density.
Notice that in the long wavelength limit, additional spacetime gradient terms are suppressed by factor of  $(q\xi_c)^n\sim (\hbar v_fq/\Delta)^n$ \cite{Altland2010}, where $\xi_c$ is the coherence length,
and $q$ can be either $|\vec{q}|$ or $|\Omega|/v_f$.
For simplicity, consider the term $\partial_t \vec{A}\cdot \nabla\times\vec{A}$ (although this term vanishes as discussed in Eq. (\ref{eq:d1_2}), it works as an illustration for the other non-axion terms, regarding the order of the coefficients).
The coefficient $C$ of this term should be on order of $\sim \frac{1}{\Delta}(\frac{v_f}{\Delta})\frac{n_s}{2m}$.
Using $n_s\sim k_f^3$, it is straightforward to obtain $C\sim (\epsilon_f/\Delta)^2$.
On the other hand, recall that the coefficients of the axion terms are of $O[(\Delta/\epsilon_f)^0]$.
Therefore, generically, the non-axion terms can be much larger, i.e., enhanced by a factor of $(\epsilon_f/\Delta)^2$ compared with the axion terms.
However, we note that none of the non-axion terms can contribute to the effects like transverse supercurrent as discussed in Sec. \ref{sec:transverse_current},
and in fact, they do not exhibit magnetoelectric effects.


\end{document}